\documentclass[12pt]{article}
\usepackage{fullpage,makeidx}
\usepackage{epsfig,color,amsmath,epstopdf}
\usepackage{textcomp,amssymb,bm}
\usepackage{graphicx,caption}
\usepackage{cancel,times,cite}

\title{Fluctuation theorems for excess and housekeeping heats for underdamped systems}

\author{Sourabh Lahiri$^1$\footnote{Email: lahiri@kias.re.kr} \hspace{0.2cm} and A. M. Jayannavar$^2$}

\date{}
\numberwithin{equation}{section} 

\makeindex

\usepackage[colorlinks,linkcolor={black},citecolor={blue}]{hyperref}

\begin{document}

\maketitle{}

\newcommand{\nwc}{\newcommand}
\nwc{\la}{\langle}
\nwc{\ra}{\rangle}
\nwc{\nn}{\nonumber}
\nwc{\Ra}{\Rightarrow}
\nwc{\wt}{\widetilde}
\nwc{\td}{\tilde}
\nwc{\lw}{\linewidth}
\nwc{\dg}{\dagger}
\nwc{\mL}{\mathcal{L}}

\nwc{\Tr}[1]{\underset{#1}{\mbox{Tr}}~}
\nwc{\av}[1]{\left< #1\right>}
\nwc{\pd}[2]{\frac{\partial #1}{\partial #2}}
\nwc{\ppd}[2]{\frac{\partial^2 #1}{\partial #2^2}}

\nwc{\zprl}[3]{Phys. Rev. Lett. ~{\bf #1},~#2~(#3)}
\nwc{\zpre}[3]{Phys. Rev. E ~{\bf #1},~#2~(#3)}
\nwc{\zpra}[3]{Phys. Rev. A ~{\bf #1},~#2~(#3)}
\nwc{\zjsm}[2]{J. Stat. Mech. ~#1~(#2)}
\nwc{\zepjb}[3]{Eur. Phys. J. B ~{\bf #1},~#2~(#3)}
\nwc{\zrmp}[3]{Rev. Mod. Phys. ~{\bf #1},~#2~(#3)}
\nwc{\zepl}[3]{Europhys. Lett. ~{\bf #1},~#2~(#3)}
\nwc{\zjsp}[3]{J. Stat. Phys. ~{\bf #1},~#2~(#3)}
\nwc{\zptps}[3]{Prog. Theor. Phys. Suppl. ~{\bf #1},~#2~(#3)}
\nwc{\zpt}[3]{Physics Today ~{\bf #1},~#2~(#3)}
\nwc{\zap}[3]{Adv. Phys. ~{\bf #1},~#2~(#3)}
\nwc{\zjpcm}[3]{J. Phys. Condens. Matter ~{\bf #1},~#2~(#3)}
\nwc{\zjpa}[3]{J. Phys. A: Math. Gen. ~{\bf #1},~#2~(#3)}





\begin{center}
\small
$^1$ Korea Institute for Advanced Study, 85 Hoegiro, Dongdaemun-gu, Seoul 130-722, Republic of Korea 

 \vspace{0.5cm}
$^2$ Institute of Physics, Sachivalaya Marg, Bhubaneswar 751005, India
\end{center}
\normalsize

\begin{abstract}
We present a simple derivation of the integral fluctuation theorems for excess housekeeping heat for an underdamped Langevin system, without using the concept of dual dynamics. In conformity with the earlier results, we find that the fluctuation theorem for housekeeping heat holds when the steady state distributions are symmetric in velocity, whereas there is no such requirement for the excess heat. We first prove the integral fluctuation theorem for the excess heat, and then show that it naturally leads to the integral fluctuation theorem for housekeeping heat. We also derive the modified detailed fluctuation theorems for the excess and housekeeping heats.
\end{abstract}

\section{Introduction}

The second law of thermodynamics states that, during any thermodynamic process,  the total change in entropy of the universe (the system and the environment with which it interacts) never decreases with time \cite{lan}: $\Delta S_{tot}\ge 0$. For small systems, the entropy change is a highly fluctuating quantity, due to the dominance of thermal fluctuations. To arrive at the second law for such systems, we first need to extend the definition of macroscopic entropy to the level of a single phase space trajectory. Such trajectory-dependent entropy changes have been defined by Seifert \cite{sei05,sei08,sei12}.
Accordingly, one defines the trajectory-dependent changes in entropy of the  system ($\Delta s$), and of the medium or heat bath ($\Delta s_m$) in which the system is present . 
 If the heat bath is large enough so that its temperature $T$ can be assumed to stay constant with time, then $\Delta s_m = \beta Q$, where $Q$ is the heat dissipated into the bath during the process, and $\beta=1/T$ is its inverse temperature (we will set the Bolzmann constant equal to unity for convenience). The system entropy is defined as the negative logarithm of the probability distribution of the system state: $s(x,t)=-\ln p(x,t)$, and accordingly the change in system entropy from time $t=0$ to $t=\tau$ is given by
\[
\Delta s = \ln\frac{p(x(0),0)}{p(x(\tau),\tau)}.
\]
Finally, the change in total entropy along a trajectory is given by
\begin{align}
\Delta s_{tot} = \Delta s_m+\Delta s.
\end{align}

One can now derive the second law in the form $\av{\Delta s_{tot}}\ge 0$.
The angular brackets imply ensemble averaging, i.e. the experiment has been performed a large number of times and the value of total entropy change has been averaged over all such realizations of the experiment. In fact, this inequality is obtained as a corollary from the exact fluctuation theorem (FT) for total entropy \cite{sei05,sei08,sei12}, given by $\av{e^{-\Delta s_{tot}}}=1$. Here, the total entropy change is simply the summation of entropy changes of the medium or heat bath ($\Delta s_m$) and of the system ($\Delta s$): 
\begin{align}
\Delta s_{tot} = \Delta s_m+\Delta s.
\end{align}
%

In a nonequilibrium steady state or NESS, the system remains in a stationary state that is out of equilibrium.  In this case, the above inequality for second law  turns out to be very weak, because heat is always dissipated into the medium (in order to maintain the steady state) even when there is no external perturbation. To get a meaningful inequality, Oono and Paniconi \cite{oon98} had suggested the division of total heat into two parts: the housekeeping heat $Q_{hk}$ and the excess heat $Q_{ex}$. We then have, $Q=Q_{hk}+Q_{ex}$. Here, $Q_{hk}$ is the heat that is dissipated into the heat bath in order to maintain a steady state, even when the protocol does not change with time, or when it changes adiabatically (system always remains close to a steady state).  If we remove this contribution, then we are left with $Q_{ex}$ that provides a stronger inequality than the conventional second law. This is what Hatano and Sasa \cite{hat01} refer to as the modified second law inequality for transitions between steady states. If the system is initially in a steady state and is perturbed thereafter by an external time-dependent control parameter $\lambda(t)$, the authors obtain an exact fluctuation theorem for the excess heat: 
\begin{align}
\av{e^{-\beta Q_{ex}-\Delta\phi}}=1,
\end{align}
 where $\phi(x;\lambda)=-\ln p_s(x;\lambda)$ is the negative logarithm of the steady state distribution. The modified second law inequality, $\beta\av{Q_{ex}}+\av{\Delta\phi} \ge 0$, is then readily obtained by application of the Jensen's inequality to the above fluctuation theorem.

In a separate work, the integral fluctuation theorem (IFT) for the housekeeping heat was also proved by Speck and Seifert \cite{spe05}, for an overdamped particle in presence of a non-conservative force. The theorem reads
\begin{align}
\av{e^{-\beta Q_{hk}}} = 1.
\end{align}
As elaborated in \cite{gar12}, these IFTs can be derived from the corresponding detailed fluctuation theorems (DFTs), where the path ratios in the original dynamics and the so-called dual dynamics have to be invoked.

However, until recently, all works in this area were in the overdamped limit, where the system state was defined by its position only \cite{hat01,spe05,esp10,esp10a,esp10b}. 
The extension of these theorems to the case of an underdamped system has recently attracted much interest \cite{spi12,spi12a,spi12b,par13,sas13}. It has been shown that although the excess heat continues to follow an IFT (known as the Hatano-Sasa relation), the full housekeeping heat in general does not. The IFT for $Q_{hk}$ holds only when the stationary distributions are symmetric with respect to the velocity variables. To state this fact mathematically, we define $p_s(x,v;\lambda)$ is the steady state distribution at a fixed value $\lambda$ of the external perturbation. Then, for the validity of the IFT for $Q_{hk}$, we must have $p_s(x,v;\lambda) = p_s(x,-v;\lambda)$. Such a requirement is not needed for obtaining the IFT for $Q_{ex}$.

To prove these theorems in the underdamped limit, usually the dual dynamics is defined. This a generalization of the concept of detailed balance, to the cases where the dynamics violates detailed balance \cite{hat01,gar12,jar06a}.
However, more than one definition of dual dynamics has been proposed in the literature \cite{spi12,par13,sas13}, and there seems to be no universally accepted definition as yet.

In this work, we arrive at the same conclusions, using the Langevin equation for an underdamped particle, using a simple approach that does not deal with the concept of dual dynamics. We observe that for velocity-symmetric steady state distributions, the satisfaction of Hatano-Sasa identity naturally leads to the IFT for housekeeping heat. We verify that the expressions for $Q_{ex}$ and $Q_{hk}$ reduce to the known expressions for the overdamped case.  We also provide modified detailed fluctuation theorems for $Q_{ex}$ and $Q_{hk}$.

\section{Fluctuation theorem for excess heat}

\label{sec:HSI}

We consider a system that is initially in a steady state corresponding to the control parameter $\lambda_0$, with the initial distribution given by $p_s( x_0,  v_0;\lambda_0)$. Thereafter, the protocol is changed as a function of time. In addition, a non-conservative force is present that does not allow the system to equilibrate, even when the external perturbation does not change with time. In such a case, the system will relax to a NESS, corresponding to the particular value of the external protocol. For such systems, Hatano and Sasa had proved the fluctuation theorem for excess heat by simply considering an identity \cite{hat01}, which for an underdamped system would be given by
\begin{align}
\left< \prod_{k=0}^{N-1} \frac{p_s( y_{k+1}; \lambda_{k+1})}{p_s( y_{k+1}; \lambda_{k})}\right> = 1,
\label{eq:HS}
\end{align}
where $y_k=(x_k,v_k)$  at a given time instant $t=t_k$, and $k$ represents time discretization. We will assume that the initial time is $t_0=0$ and the final value of time is $t_N=\tau$. The angular brackets represent the averages taken over all trajectories in phase space, and the subscript $s$ implies steady state distributions. 
Note that in the following, we are going to adhere to the Stratonovich discretization scheme, in which the normal rules of calculus are applicable. 

Let $K(y_{k+1}|y_k;\lambda_k)$, be the transition probability from the state $(y_k)$ at time $t_k$ to the state $(y_{k+1})$ at time $t_{k+1}$, when the protocol value is fixed at $\lambda_k$.
When written  explicitly, the LHS is
\begin{align}
\int {dy}_0\cdots {dy}_N ~p_s( y_0;\lambda_0)\prod_{k=0}^{N-1}K(y_{k+1}|y_k;\lambda_k) \frac{p_s( y_{k+1}; \lambda_{k+1})}{p_s( y_{k+1}; \lambda_{k})}.
\end{align}
The integration is over full path $\{x_0,x_1,\cdots,x_N,v_1,v_2,\cdots,v_N\}$.
It can be readily shown to be equal to unity, by repeatedly using  the following property of steady state systems:
\begin{align}
\int dy_k~ K(y_{k+1}|y_k;\lambda_k)p_s(y_{k};\lambda_{k}) = p_s(y_{k+1}; \lambda_{k}).
\label{eq:ss_property}
\end{align}
Now we define an effective potential $\phi(x,v;\lambda(t))$ such that $p_s\equiv e^{-\phi}$ for steady state distributions. From eq. \eqref{eq:HS}, we  then get
\begin{align}
\left<\exp\left[-\int dt ~\dot\lambda\pd{\phi}{\lambda}\right]\right> =1 =
 \left<e^{-\Delta\phi}\exp\left[\int dt~\left(\dot x \pd{\phi}{x}+\dot v\pd{\phi}{v}\right)\right]\right>.
\end{align}
Here we have used the chain rule (see appendix \ref{sec:appA})
\begin{align}
\Delta\phi = \int dt \left[\left(\dot x \pd{\phi}{x}+\dot v\pd{\phi}{v}\right)+\dot\lambda\pd{\phi}{\lambda}\right].
\label{eq:Delta_phi}
\end{align}
Defining the excess heat as
\begin{align}
\beta Q_{ex} &= -\int dt~\left(\dot x\pd{\phi}{x}+\dot v\pd{\phi}{v}\right),
\label{eq:Qex}
\end{align}
we get the fluctuation theorem for $Q_{ex}$:
\begin{align}
\la e^{-\beta Q_{ex}-\Delta \phi}\ra = 1.
\label{eq:HS_gen}
\end{align}
The above definition of the excess heat is a simple generalization of the one in the overdamped regime, where the second term in the integrand of \eqref{eq:Qex} will be absent. The non-adiabatic entropy is defined as
\begin{align}
\Delta s_{na} = \beta Q_{ex}+\Delta \phi.
\end{align}
Thus eq. \eqref{eq:HS_gen} can also be written as 
\begin{align}
\av{e^{-\Delta s_{na}}}=1.
\end{align}
It is reasonable to call this quantity  the ``non-adiabatic entropy'' because it vanishes in an adiabatic process (i.e., when the system driven slowly enough, so that it is always in the steady state distribution corresponding to the instantaneous value of the protocol):
\begin{align}
\av{\Delta s_{na}} =& \av{\int dt ~\dot{\lambda}\pd{\phi}{\lambda}}\nn\\
 =& \int dt~\dot{\lambda}\int dxdv ~e^{-\phi(x,v;\lambda)} \pd{\phi}{\lambda} \nn\\
 =& -\int dt~\dot{\lambda}\pd{}{\lambda}\int dxdv ~e^{-\phi(x,v;\lambda)}=0,
\end{align}
due to normalization of steady state distribution.

Note that the derivation of \eqref{eq:HS_gen} does not use any information about the time-reversed trajectories, and hence is valid irrespective whether or not the steady state distribution is even in velocity. 

\section{General case: arbitrary initial distributions}

We have proved the IFT for $Q_{ex}$ for the case where the system begins in  a steady state. To verify its validity for the general case of an arbitrary initial distribution, on needs to begin with the generalized version of eq. \eqref{eq:HS}, which is given by
\begin{align}
\left< A(\tau)\prod_{k=0}^{N-1} \frac{p_s( y_{k+1}; \lambda_{k+1})}{p_s( y_{k+1}; \lambda_{k})}\right> = \int dx_Ndv_N A(\tau)p_s(y_N;\lambda_N) =  \av{A(\tau)}_{p_s(\tau)},
\end{align}
where $\tau=t_N$, and the RHS is simply an average with respect to the steady state distribution corresponding to the final value of the protocol. $A(\tau)$ is the compact form for $A(y_N(\tau))$. The relation can be proved by application of the property \eqref{eq:ss_property}. Note that substituting $A(\tau)=1$ readily leads to the relation \eqref{eq:HS}. 

Now, let $p(y_N;\lambda_N)$ be an arbitrary normalized distribution of the final states. Substituting $A(\tau)=p(y_N;\lambda_N)/p_s(y_N;\lambda_N)$, we find that the RHS is once again equal to unity:
\begin{align}
\left< \frac{p(y_N;\lambda_N)}{p_s(y_N;\lambda_N)}\prod_{k=0}^{N-1} \frac{p_s( y_{k+1}; \lambda_{k+1})}{p_s( y_{k+1}; \lambda_{k})}\right> = \int dx_Ndv_N \frac{p(y_N;\lambda_N)}{p_s(y_N;\lambda_N)}p_s(y_N;\lambda_N) =  1.
\label{eq:A}
\end{align}
Since (see sec. \ref{sec:HSI})
\[
\prod_{k=0}^{N-1} \frac{p_s( y_{k+1}; \lambda_{k+1})}{p_s( y_{k+1}; \lambda_{k})} = e^{-\beta Q_{ex}-\Delta\phi}, 
\]
we can rewrite eq. \eqref{eq:A} as
\begin{align}
\av{\frac{p(y_N;\lambda_N)}{p_s(y_N;\lambda_N)}e^{-\beta Q_{ex}-\Delta\phi}}=1.
\end{align}
Writing the LHS explicitly, we get,
\begin{align}
1 &=\int {dy}_0\cdots {dy}_N ~e^{-\beta Q_{ex}-\Delta\phi}\frac{p(y_N;\lambda_N)}{p_s(y_N;\lambda_N)}p_s( y_0;\lambda_0)\prod_{k=0}^{N-1}K(y_{k+1}|y_k;\lambda_k)  \nn\\
&= \int {dy}_0\cdots {dy}_N ~e^{-\beta Q_{ex}-\Delta\phi}\frac{p(y_N;\lambda_N)}{p_s(y_N;\lambda_N)}\frac{p_s( y_0;\lambda_0)}{p( y_0;\lambda_0)}p( y_0;\lambda_0)\prod_{k=0}^{N-1}K(y_{k+1}|y_k;\lambda_k)  \nn\\
&= \int {dy}_0\cdots {dy}_N ~e^{-\beta Q_{ex}-\Delta\phi} e^{\Delta\phi-\Delta s} p( y_0;\lambda_0)\prod_{k=0}^{N-1}K(y_{k+1}|y_k;\lambda_k) \nn\\
&=\av{e^{-\beta Q_{ex}-\Delta s}}.
\end{align}
In the second line, we have multiplied and divided $p_s (y_0;\lambda_0)$ by an arbitrary initial distribution $p( y_0;\lambda_0)$. In the next step, the definition $\Delta s = \ln[p( y_0;\lambda_0)/p( y_N;\lambda_N)]$ has been used. Thus, the IFT for excess heat is given in the most general form by
\begin{align}
\av{e^{-\beta Q_{ex}-\Delta s}}=1.
\label{eq:HS_arb}
\end{align}

\section{Fluctuation theorem for housekeeping heat}

As mentioned earlier, the  housekeeping heat $Q_{hk}$ is just the difference between the total dissipated heat $Q$ and the excess heat $Q_{ex}$: 
\begin{align}
Q_{hk} &= Q-Q_{ex}.
\end{align}
To prove the FT for $Q_{hk}$, we first write down a concrete Langevin equation that includes a non-conservative force $f_{nc}(t)$ and a conservative force $f_c(t)$. If the total force is given by $ft)=f_{nc}+f_c$, then
\begin{align}
m\dot v = -\gamma v + f(t) + \xi(t).
\end{align}
Let us denote the full trajectory of the system in the phase space by $(X(t),V(t))$. The probability of a forward trajectory $P_+ \equiv P[X(t),V(t)|x_0,v_0]$, from the initial point $(x_0,v_0)$, is given by \cite{dha04}
\begin{align}
P_+ \sim \exp\left[-\frac{\beta}{4\gamma}\int_0^\tau dt(m\dot v+\gamma v -f)^2\right].
\label{eq:P+}
\end{align}
We will consider the case when the system begins from an initial steady state, and also ends in a final steady state that is in general different from the initial one. We will further assume that the force $f(t)$ \emph{ is independent of velocities}.

To generate the reverse process, we change the time-dependence of $f(t)$ to $f(\tau-t)$.
The reverse trajectory $(X^R(t),V^R(t))$ corresponding to the forward trajectory $(X(t),V(t))$ is defined as the one in which the variable $(x,v)$ at a given time instant changes to $(x,-v)$, while the sequence of state transitions is reversed. In other words, if the forward trajectory is given by
$(x_0,v_0)\to (x_1,v_1)\to \cdots \to (x_{N-1},v_{N-1})\to (x_N,v_N)$, then the reverse trajectory will consist of the sequence of states $(x_N,-v_N)\to (x_{N-1},-v_{N-1})\to \cdots \to (x_1,-v_1)\to (x_0,-v_0)$.

Next, by switching the sign of velocity, we can write the probability for the reverse trajectory, $P_- \equiv P[X^R(t),V^R(t)|x_\tau,v_\tau]$, as
\begin{align}
P_- \sim \exp\left[-\frac{\beta}{4\gamma}\int_0^\tau dt(m\dot v-\gamma v -f)^2\right].
\label{P_-}
\end{align}
The normalization constants being the same, the ratio of the trajectories is given by
\begin{align}
\frac{P_+}{P_-} = \exp\left[-\beta\int_0^\tau dt ~v(m\dot v-f)\right] = \exp\left[\beta\int_0^\tau dt ~ v (\gamma v-\xi)\right].
\end{align}
The multiplications are of Stratonovich type, so that we must use $v=[v(t)+v(t+\Delta t)]/2$, where $\Delta t$ is the time step for discretization.
We know, from stochastic thermodynamics, that the above quantity is simply the heat $Q$ dissipated into the bath by the system, along the forward trajectory \cite{sek97,sek98,cro98,cro00}. 

Now, we \emph{choose} the initial distribution of the reverse process as the  steady state distribution corresponding to the final value $\lambda_\tau$ of the protocol in the forward process. In other words, the initial states of the reverse process are sampled from $p_s^\tau(x_\tau,v_\tau;\lambda_\tau)$. This is in general a different distribution from $p_s^\tau(x_\tau,-v_\tau;\lambda_\tau)$, the latter being the time-reversed distribution of $p_s^\tau(x_\tau,v_\tau;\lambda_\tau)$. This particular choice  used by us provides a clear physical meaning to the ratio of forward to reverse trajectories, namely the change in total entropy for a system beginning and ending in steady states, even when the steady state distribution is asymmetric in velocity.

We now multiply the ratio $P_+/P_-$ by the initial distributions to obtain
\begin{align}
\frac{P_+ ~p_s^0(x_0,v_0)}{P_- ~p_s^\tau(x_\tau,v_\tau)} = e^{\beta Q+\Delta \phi}.
\label{eq:ratio}
\end{align}
If the system begins and ends in (nonequilibrium) steady states, then the total entropy change of the system and heat bath during the process is given by 
\begin{align}
\Delta s_{tot} = \beta Q+\Delta\phi.
\end{align}
This is the quantity that appears in the exponent of the right hand side of \eqref{eq:ratio}.

Next, we use the relation $Q=Q_{ex}+Q_{hk}$, to rewrite the above equation in the form
\begin{align}
\frac{P_+ ~p^s_0(x_0,v_0)}{P_- ~p^s_\tau(x_\tau,v_\tau)} = e^{\beta(Q_{ex}+Q_{hk})+\Delta s}.
\label{eq:DFT1}
\end{align}
A simple cross-multiplication gives
\begin{align}
\av{e^{-\beta Q_{ex}-\Delta s}}_F = \av{e^{\beta Q_{hk}}}_R        
= 1.
\label{eq:cross_mult}
\end{align}
Till now, no assumption has been made on the form of stationary distribution.
Now, we consider the case when the steady state distribution is \emph{even} in velocity, so that we have $\phi(x,v;\lambda)=\phi(x,-v;\lambda)$. Since 
\[
\beta Q_{ex} =- \int dt \left(\dot x\pd{\phi}{x}+\dot v\pd{\phi}{v}\right),
\label{eq:Qex}
\]
 (see sec. \ref{sec:HSI}), we  note that $Q_{ex}$ reverses sign in the steady state. 
The total heat, given by
\[
Q=\int_0^\tau dt ~ v (f-m\dot v)
\]
also changes sign under time-reversal. This implies that for velocity symmetric $\phi$, the housekeeping heat $Q_{hk}=Q-Q_{ex}$ changes sign as well, so that we can write from \eqref{eq:cross_mult}
\begin{align}
\av{e^{-\beta Q^R_{hk}}}_R =1.
\end{align}
 Now, it is up to the observer to decide which one is the forward and which is the reverse process. All integral fluctuation theorems are equally valid in either process, so that we can as well write
\begin{align}
\la{e^{-\beta Q_{hk}}}\ra_F=1.
\end{align}
This is the integral fluctuation relation for the housekeeping heat.

The explicit expression for $Q_{hk }$ can be written down from its  definition:
\begin{align}
Q_{hk} &= Q-Q_{ex} \nn\\
&= \int_0^\tau dt ~v(f-m\dot v) +T\int dt~\left(\dot x\pd{\phi}{x}+\dot v\pd{\phi}{v}\right).
\end{align}
When steady state distributions are even in velocity, both $Q$ and $Q_{ex}$ switch signs under time-reversal, and so does $Q_{hk}$. We can readily find that the above definition reduces to the expression for $Q_{hk}$ in the overdamped case \cite{gar12}. In this regime, neglecting the $\dot v$ terms in the above definition, we get
\begin{align}
Q^{(ov)}_{hk} &= \int_0^\tau dt ~\dot x \left(f+T\pd{\phi}{x}\right) = \gamma \int_0^\tau dt ~\dot x v_{s},
\end{align}
$v_s$ being the local velocity in the steady state. This verification acts as a consistency check on the expression for excess and housekeeping heats obtained for underdamped systems.

Finally, we note that the IFT for housekeeping heat breaks down if $\phi$ is asymmetric in velocity.

\section{The modified detailed fluctuation theorems}

We now follow the approach of \cite{han09c,noh12}, in order to derive the modified detailed fluctuation theorems for the excess and the housekeeping heats. Once again, we assume that the stationary state distribution is even under time-reversal: $\phi(x,v;\lambda)=\phi(x,-v;\lambda)$. Then eq. \eqref{eq:DFT1} can be converted to the following form:
\begin{align}
\frac{P_f(Q_{ex},Q_{hk},\Delta\phi)}{P_r(-Q_{ex},-Q_{hk},-\Delta\phi)} = e^{\beta(Q_{ex}+Q_{hk})+\Delta\phi},
\label{eq:DFT2}
\end{align}
where 
\[
P_f(Q_{ex},Q_{hk},\Delta\phi) = \av{\delta(\mathcal Q_{ex}[X,V]-Q_{ex})\delta(\mathcal Q_{hk}[X,V]-Q_{hk})\delta(\Phi[X,V]-\Delta\phi)},
\]
and
\[
P_r(Q_{ex},Q_{hk},\Delta\phi) = \av{\delta(\mathcal Q_{ex}[\td X,\td V]+Q_{ex})\delta(\mathcal Q_{hk}[\td X,\td V]+Q_{hk})\delta(\Phi[\td X,\td V]+\Delta\phi)}.
 \]
 Here, $\mathcal Q_{ex}$, $\mathcal Q_{hk}$ and $\Phi$ are functions of the trajectory $(X,V)$. The subscripts $f$ and $r$ denote the forward and the reverse processes, respectively.
 
 The above DFT (eq.\eqref{eq:DFT2}) for the joint probability distributions can be rewritten in several forms. For instance, the FT for $Q_{ex}$ can be written down as follows:
\begin{align}
\int dQ_{hk}d\Delta\phi ~P_r(-Q_{ex},-Q_{hk},-\Delta\phi) =& e^{-\beta Q_{ex}}\int dQ_{hk}d\Delta\phi ~P_f(Q_{ex},Q_{hk},\Delta\phi)e^{-\beta Q_{hk}-\Delta\phi} \nn\\
\Ra P_r(-Q_{ex}) =& e^{-\beta Q_{ex}}P_f(Q_{ex})\int dQ_{hk}d\Delta\phi ~P_f(Q_{hk},\Delta\phi|Q_{ex})e^{-\beta Q_{hk}-\Delta\phi} \nn\\
\Ra \frac{P_f(Q_{ex})}{P_r(-Q_{ex})} =& \frac{e^{\beta Q_{ex}}}{\Psi(Q_{ex})},
\label{MDFT}
\end{align}
where
\begin{align}
\Psi(Q_{ex}) = \int dQ_{hk}d\Delta\phi ~P_f(Q_{hk},\Delta\phi|Q_{ex})e^{-\beta Q_{hk}-\Delta\phi} \equiv \av{e^{-\beta Q_{hk}-\Delta\phi}|Q_{ex}}.
\end{align}
Eq. \eqref{MDFT} is the modified detailed fluctuation theorem for the excess heat (usually, the conventional form of DFT for a variable $\Sigma$ is given by $\frac{P_f(\Sigma)}{P_r(-\Sigma)}=e^{\Sigma}$).
Similarly, one can derive
\begin{align}
\frac{P_f(Q_{hk})}{P_r(-Q_{hk})} = \frac{e^{\beta Q_{hk}}}{\Psi(Q_{hk})},
\end{align}
where
\begin{align}
\Psi(Q_{hk}) = \av{e^{-\beta Q_{ex}-\Delta\phi}|Q_{hk}}.
\end{align}
In terms of entropies,  eq. \eqref{eq:DFT2} can be put into the form
\begin{align}
\frac{P_f(\Delta s_a,\Delta s_{na})}{P_r(-\Delta s_a,-\Delta s_{na})} = e^{\Delta s_a+\Delta s_{na}},
\end{align}
where $\Delta s_a = \beta Q_{hk}$ and $\Delta s_{na}=\beta Q_{ex}+\Delta\phi$ are the  adiabatic and non-adiabatic entropy changes, respectively \cite{gar12}.

\section{Conclusions}

In this work, we have derived the integral fluctuation relations for the excess and housekeeping heats for underdamped Langevin systems, in a simplistic way, without defining the dual dynamics. We have found that $Q_{ex}$ always follows the integral fluctuation theorem, irrespective of the presence of velocity variables that switch sign under time-reversal. However, $Q_{hk}$ follows an IFT only if the steady state distributions are even in velocity. We have shown that  the definitions of both these quantities reduce to the definitions in overdamped regime, when the inertia terms are neglected. The modified detailed fluctuation theorems have been provided, following the approach of \cite{han09c,noh12}. We believe that this approach will help in simplifying the understanding of excess and housekeeping heats in underdamped systems.

\section{Acknowledgement}

AMJ thanks DST, India for financial support. SL thanks Prof. Hyunggyu Park    for useful discussions.

\appendix

\section{Chain rule for derivative}
\label{sec:appA}

Note that , since $\Delta v\sim \sqrt{\Delta t}$ for the underdamped system, the Taylor expansion should ideally have been
\begin{align}
d\phi \simeq \pd{\phi}{\lambda}\Delta\lambda + \pd{\phi}{x}\Delta x + \pd{\phi}{v}\Delta v + \ppd{\phi}{v}\frac{\Delta v^2}{2}.
\label{chainrule}
\end{align}
However, in the Stratonovich scheme, we must use $\bar{v}=v+\Delta v/2$ in the argument of $\phi$, so that
\begin{align}
\phi(\bar{v}) &= \phi(v+\Delta v/2) = \phi(v)+\frac{\Delta v}{2}\pd{\phi(v)}{v} \nn\\
\Ra \pd{\phi(\bar v)}{v}\Delta v &=  \pd{\phi(v)}{v}\Delta v+\ppd{\phi}{v}\frac{\Delta v^2}{2}.
\end{align}
Thus, if $\phi(x,v;\lambda)$ is interpreted in the Stratonovich sense,
then we can simply write
\begin{align}
d\phi \simeq \pd{\phi}{\lambda}\Delta\lambda + \pd{\phi}{x}\Delta x + \pd{\phi}{v}\Delta v.
\end{align}
As a result, we do not need to retain the second derivative terms in velocity.

\end{document}